\documentclass[12pt]{article}

\usepackage[dvipdfmx]{graphicx}
\usepackage{subcaption}
\usepackage{amsmath,amssymb}
\usepackage{bm}
\usepackage{graphicx, color}
\usepackage{wrapfig}
\usepackage{cite}
\usepackage{braket}
\usepackage{float}
\begin{document}
\title{
\ \\*[-30pt]
\normalsize
{\Large \bf
Flavino Dark Matter\\ in a Non-Abelian Discrete Flavor Model
\\*[20pt]}}

\author{
\centerline{
Takaaki~Nomura $^{1}$\footnote{nomura@scu.edu.cn},~
Yusuke~Shimizu $^{2,3}$\footnote{shimizu.yusuke@kaishi-pu.ac.jp}, and
~Towa~Takahashi $^{2}$\footnote{towa@muse.sc.niigata-u.ac.jp}}
\\*[20pt]
\centerline{
\begin{minipage}{\linewidth}
\begin{center}
$^1${\it \normalsize
College~of~Physics,~Sichuan~University,~Chengdu~610065,~China
} \\*[5pt]
$^2${\it \normalsize
Graduate~School~of~Science~and~Technology,~Niigata~University, \\
Niigata,~950-2181,~Japan
} \\*[5pt]
$^3${\it \normalsize 
Department~of~Information,~Kaishi~Professional~University, \\
Niigata~950-0916,~Japan
}
\end{center}
\end{minipage}
}
\\*[70pt]}

\date{
\centerline{\small \bf Abstract}
\begin{minipage}{0.9\linewidth}
\medskip
\medskip
\small
We study a relic density of the ``flavino'' dark matter in modified Altarelli and Feruglio $A_4$ model 
which is respecting the $SU(2)_L\times A_4\times Z_3\times U(1)_R$ symmetry. 
We calculate the Lagrangian from the superpotential in the model. 
In estimating the relic density, we consider the relevant interactions from the Lagrangian 
that realize the vacuum expectation value alignments and charged lepton masses 
where we assume that the supersymmetry breaking effects 
are small for ``flavon'' sector.  
As a result, we find the degenerate masses between the lightest ``flavon'' and ``flavino'', 
and only two parameters in the potential determines the relic density.
Then the allowed parameter space of these parameters are estimated from the relic density calculation taking a 
constraint from lepton flavor violation into account. 
We also briefly discuss other dark matter physics such as the direct detection, indirect detection, and collider search.
\end{minipage}
}

\begin{titlepage}
\maketitle
\thispagestyle{empty}
\end{titlepage}
\newpage
\section{Introduction}
The standard model (SM) is one of the successful ones with discovery of the Higgs boson. 
However, there are some mysterious problems in the SM, e.g. the origin of the generations for particle physics. 
In the SM, one cannot explain the differences of the mixing angles and masses for the quarks and leptons. 
Yukawa couplings are completely free parameters so that the mixing angles and masses cannot be predicted 
in the SM. Furthermore, the neutrinos are massless for the renormalizable operators.
One of the attractive phenomena to solve the problems is the neutrino oscillation 
which is the evidence of the beyond the SM. 
Actually, the neutrino oscillation experiments provide us the important informations which are 
the two neutrino mass squared differences and two large mixing angles. The reactor neutrino experiments 
reported the last mixing angle which is non-zero value~\cite{DayaBay:2012fng,RENO:2012mkc}. 
Furthermore, the neutrino oscillation experiments go to new stage to measure the Dirac CP phase in the lepton sector. 
The neutrinos are thus windows to the new physics beyond the SM. 

The flavor symmetry can be applied to the three generations of the SM fermions. 
The Froggatt-Nielsen mechanism~\cite{Froggatt:1978nt} which was proposed 
by C. D.  Froggatt and H. B. Nielsen applies a global $U(1)_{FN}$ symmetry. 
Thanks to the symmetry, one can naturally explain the fermion mass hierarchies. 
The non-Abelian discrete symmetry (See for the review \cite{Altarelli:2010gt}-\cite{Kobayashi:2022moq}.) 
can naturally explain the lepton mixing angles which are 
tri-bimaximal mixing (TBM)~\cite{Harrison:2002er,Harrison:2002kp} before the reactor experiments 
reported the non-zero reactor angle $\theta _{13}$~\cite{DayaBay:2012fng,RENO:2012mkc}. 
After measurements of the $\theta _{13}$, many authors have studied the breaking or deviation 
from the TBM~\cite{Xing:2002sw}-\cite{King:2013eh} or other patterns of the lepton mixing angles, 
e.g. tri-bimaximal-Cabibbo mixing~\cite{King:2012vj,Shimizu:2012ry}. 
One of the successful flavor models was proposed 
by G. Altarelli and F. Feruglio~\cite{Altarelli:2005yp,Altarelli:2005yx}. 
The Altarelli and Feruglio (AF) model realizes the TBM by using the non-Abelian discrete symmetry $A_4$. 
They mainly introduced two SM gauge singlet scalar fields so-called “flavons”. Taking 
the vacuum expectation value (VEV) alignments of the $A_4$ triplets as $(1, 0, 0)$ and $(1, 1, 1)$, 
one can naturally explain that the charged lepton mass matrix is diagonal and neutrino mixing is TBM, respectively. 
However, the correct VEV alignments cannot be driven from the potential analysis. 
Then, they adopted the supersymmetry (SUSY) and introduced 
the so-called “driving” fields~\cite{Altarelli:2005yp,Altarelli:2005yx}. 

Introduction of the SUSY can solve another mysterious problem that is existence of dark matter (DM) in the universe
since the lightest supersymmetric particle (LSP) is stable due to $R$-parity and it can be DM if it is neutral one~\cite{Farrar:1978xj, Ellis:1983ew}.
In the minimal supersimmetric SM (MSSM), the lightest neutralino is a good DM candidate and its properties are well discussed 
such as parameter space explaining observed relic density~\cite{Roszkowski:2017nbc}.
Also the neutralino DM is searched for in direct/indirect detection and collider experiments~\cite{Baer:2004qq,Baer:2005bu, Han:2014nba, GAMBIT:2018gjo, ATLAS:2021moa, CMS:2021few}.
However we have not seen any clear evidence of a DM in these experiments yet and nature of DM is an open question.
Interestingly, in SUSY models with flavons, we have new candidate of DM that is the superpartner of the flavon which we call the "flavino"~\cite{Chigusa:2018yua}.
The lightest flavino can be DM if it is the LSP since flavinos are neutral particles.
Thus it is worth exploring the possibility of flavino DM.
 

In this paper, we discuss a relic density of the flavino DM 
in modified AF $A_4$ model~\cite{Shimizu:2011xg},\cite{Muramatsu:2016bda}-\cite{Muramatsu:2017xmn} 
which is respecting the $SU(2)_L\times A_4\times Z_3\times U(1)_R$ symmetry. 
In estimating the relic density, we consider relevant interactions from superpotential that realize the VEV alignments and charged lepton masses
where we assume SUSY breaking effects are small for flavon sector.  
As a result, we find degenerate mass between the lightest flavon and flavino, and only two parameters in the potential determines the relic density.
Then allowed parameter space of these parameters are estimated from relic density calculation taking a constraint from lepton flavor violation (LFV) into account.
We also briefly discuss other DM physics such as the direct detection, indirect detection and collider search.

This paper is organized as follows. In Section~\ref{sec:Model}, we briefly introduce the modified AF $A_4$ model 
which is respecting the $SU(2)_L\times A_4\times Z_3\times U(1)_R$ symmetry. 
In Section~\ref{sec:Dark-matter}, we discuss the relic density of the flavino DM 
calculating its annihilation cross section.
Section~\ref{sec:Summary} is devoted to summary. 
We show the relevant multiplication rule of the $A_4$ group in Appendix~\ref{sec:multiplication-rule}.
The LFV constraint for our model is shown in~Appendix~\ref{sec:LFVconstraint}.

\section{Model}
\label{sec:Model}
In this section, we briefly introduce an $SU(2)_L\times A_4\times Z_3\times U(1)_R$ 
model~\cite{Shimizu:2011xg},\cite{Muramatsu:2016bda}-\cite{Muramatsu:2017xmn}. 
Three generations of the left-handed lepton doublet superfields 
$\Phi _\ell =(\Phi _{\ell 1},\Phi _{\ell 2},\Phi _{\ell 3})$ are assigned to the triplet as $\bf 3$ for $A_4$ symmetry. 
The right-handed charged lepton superfields $\Phi _{e_R^c}$, $\Phi _{\mu _R^c}$, and $\Phi _{\tau _R^c}$ are 
assigned to the singlets as $\bf 1$, $\bf 1''$, and $\bf 1'$ for $A_4$ symmetry, respectively. 
One of the Higgs doublet superfields is defined as $\Phi _d$ and assigned to the singlet as 
$\bf 1$ for $A_4$ symmetry. 
We introduce superfields $\Phi _T=(\Phi _{T1},\Phi _{T2},\Phi _{T3})$ 
which are so-called ``flavon superfields'' and assigend to the triplet as $\bf 3$ 
for $A_4$ symmetry.
In order to obtain the relevant couplings, we add the $Z_3$ symmetry. 
We also add the $U(1)_R$ symmetry so that we can generate the VEVs and VEV alignments 
through the $F$-terms by coupling flavon superfields to so-called 
``driving superfields'' $\Phi _0^T=(\Phi _{01}^T,\Phi _{02}^T,\Phi _{03}^T)$ 
which are assigned to triplet as $\bf 3$ for $A_4$ symmetry 
and carry the $R$ charge $+2$ under $U(1)_R$ symmmetry.
The charge assignments of $SU(2)_L$, $A_4$, $Z_3$, and $U(1)_R$ are summarized in Tab.~\ref{tab:model}.
\begin{table}[h]
  \centering
  \begin{tabular}{|c||cccc|ccc|}
    \hline 
    \rule[14pt]{0pt}{0pt}
    & $\Phi _\ell $ & $\Phi _{e_R^c}$ &$\Phi _{\mu_R^c}$ & 
    $\Phi _{\tau_R^c}$ & 
    $\Phi _d$ & $\Phi _T$ & $\Phi _0^T$
    \\ \hline 
    \rule[14pt]{0pt}{0pt}
    $SU(2)_L$ & $2$ & $1$ & $1$ & $1$ & $2$ & $1$ & $1$
    \\
    $A_4$ & ${\bf 3}$ & ${\bf 1}$ & ${\bf 1''}$ & ${\bf 1'}$ & ${\bf 1}$ & ${\bf 3}$ & ${\bf 3}$ \\
    $Z_3$ & $\omega $ & $\omega ^2$ & $\omega^2$ & $\omega ^2$ & $1$ & $1$ & $1$ \\
    $U(1)_R$ & $1$ & $1$ & $1$ & $1$ & $0$ & $0$ & $2$ \\
    \hline
  \end{tabular}
  \caption{The charge assignments of $SU(2)_L\times A_4\times Z_3\times U(1)_R$ symmetry in our model.}
  \label{tab:model}
\end{table}
In these setup, we can now write douwn the superpotential for respecting 
$SU(2)_L\times A_4\times Z_3\times U(1)_R$ symmetry at the leading order in terms of 
the $A_4$ cut-off scale $\Lambda $ as 
\begin{align}
w_{\Phi _T}&=w_{\ell }+w_d^T, \nonumber \\
w_{\ell }&=y_e\Phi _T\Phi _\ell \Phi _{e_R^c}\Phi _d/\Lambda + y_\mu \Phi _T\Phi _\ell \Phi _{\mu _R^c}\Phi _d/\Lambda + y_\tau \Phi _T\Phi _\ell \Phi _{\tau _R^c}\Phi _d/\Lambda ,\nonumber \\
w_d^T&=-M\Phi _0^T\Phi _T+g\Phi _0^T\Phi _T\Phi _T,
\label{eq:superpotential}
\end{align}
where $\Phi _i$'s are chiral superfields $\Phi _i=\phi _i+\sqrt{2}\theta \psi _i+\theta \theta F_i$, $y_j$'s are Yukawa couplings, $M$ is mass paremeter for the flavon superfields $\Phi _T$ and $\Phi _0^T$~\footnote{Hereafter we also call deriving field $\Phi_0^T$ as flavon field for simplicity.}, and $g$ is trilinear coupling for flavon superfields. 
The superpotential in Eq.~\eqref{eq:superpotential} is rewritten as
\begin{align}
w_{\ell }&=y_e(\Phi _{T1}\Phi _{\ell _1}+\Phi _{T2}\Phi _{\ell _3}+\Phi _{T3}\Phi _{\ell _2})\Phi _{e_R^c}\Phi _d/\Lambda \nonumber \\
&+y_\mu (\Phi _{T3}\Phi _{\ell _3}+\Phi _{T1}\Phi _{\ell _2}+\Phi _{T2}\Phi _{\ell _1})\Phi _{\mu _R^c}\Phi _d/\Lambda \nonumber \\
&+y_\tau (\Phi _{T2}\Phi _{\ell _2}+\Phi _{T3}\Phi _{\ell _1}+\Phi _{T1}\Phi _{\ell _3})\Phi _{\tau _R^c}\Phi _d/\Lambda , \nonumber \\
w_d^T&=-M(\Phi _{01}^T\Phi _{T1}+\Phi _{02}^T\Phi _{T3}+\Phi _{03}^T\Phi _{T2}) \nonumber \\
&+\frac{2}{3}g\left [\Phi _{01}^T(\Phi _{T1}\Phi _{T1}-\Phi _{T2}\Phi _{T3})+\Phi _{02}^T(\Phi _{T2}\Phi _{T2}-\Phi _{T3}\Phi _{T1})+\Phi _{03}^T(\Phi _{T3}\Phi _{T3}-\Phi _{T1}\Phi _{T2})\right ].
\end{align}
Then, the Lagrangian $\mathcal {L}_{\Phi _T}$ for our model is given as
\begin{align}
\mathcal{L}_{\Phi _T}&=\int d^2\theta w_{\Phi _T}+\int d^2\bar \theta \bar w_{\Phi _T}-V_{\Phi _T}, \nonumber \\
V_{\Phi _T}&=V_\ell +V_T, 
\end{align}
where the scalar potential $V_\ell$ and $V_T$ are obtained from $w_d^T$ and $w_\ell$ respectively. 
The scalar potential $V_T$ is obtained as
\begin{align}
V_T&=\sum _X\left |\frac{\partial w_d^T}{\partial X}\right |^2 \nonumber \\
&=\left |-M\phi _{T1}+\frac{2}{3}g(\phi _{T1}\phi _{T1}-\phi _{T2}\phi _{T3})\right |^2 \nonumber \\
&+\left |-M\phi _{T3}+\frac{2}{3}g(\phi _{T2}\phi _{T2}-\phi _{T3}\phi _{T1})\right |^2 \nonumber \\
&+\left |-M\phi _{T2}+\frac{2}{3}g(\phi _{T3}\phi _{T3}-\phi _{T1}\phi _{T2})\right |^2 \nonumber \\
&+\left |-M\phi _{01}^T+\frac{2}{3}g(2\phi _{01}^T\phi _{T1}-\phi _{02}^T\phi _{T3}-\phi _{03}^T\phi _{T2})\right |^2 \nonumber \\
&+\left |-M\phi _{03}^T+\frac{2}{3}g(2\phi _{02}^T\phi _{T2}-\phi _{03}^T\phi _{T1}-\phi _{01}^T\phi _{T3})\right |^2 \nonumber \\
&+\left |-M\phi _{02}^T+\frac{2}{3}g(2\phi _{03}^T\phi _{T3}-\phi _{01}^T\phi _{T2}-\phi _{02}^T\phi _{T1})\right |^2,
\label{eq:Lagrangian}
\end{align}
where $X=\phi _{T1}$, $\phi _{T2}$, $\phi _{T3}$, $\phi _{01}^T$, $\phi _{02}^T$, and $\phi _{03}^T$. 
On the other hand, the scalar potential $V_\ell $ is written by 
\begin{align}
V_\ell &=\sum _Y\left |\frac{\partial w_\ell }{\partial Y}\right |^2 \nonumber \\
&=\left |(y_e\tilde \phi _{\ell _1}\tilde \phi _{e_R^c}+y_\mu \tilde \phi _{\ell _2}\tilde \phi _{\mu _R^c}+y_\tau \tilde \phi _{\ell _3}\tilde \phi _{\tau _R^c})h_d/\Lambda \right |^2 \nonumber \\
&+\left |(y_e\tilde \phi _{\ell _3}\tilde \phi _{e_R^c}+y_\mu \tilde \phi _{\ell _1}\tilde \phi _{\mu _R^c}+y_\tau \tilde \phi _{\ell _2}\tilde \phi _{\tau _R^c})h_d/\Lambda \right |^2 \nonumber \\
&+\left |(y_e\tilde \phi _{\ell _2}\tilde \phi _{e_R^c}+y_\mu \tilde \phi _{\ell _3}\tilde \phi _{\mu _R^c}+y_\tau \tilde \phi _{\ell _1}\tilde \phi _{\tau _R^c})h_d/\Lambda \right |^2 \nonumber \\
&+\left |(y_e\phi _{T1}\tilde \phi _{e_R^c}+y_\mu \phi _{T2}\tilde \phi _{\mu _R^c}+y_\tau \phi _{T3}\tilde \phi _{\tau _R^c})h_d/\Lambda \right |^2 \nonumber \\
&+\left |(y_e\phi _{T3}\tilde \phi _{e_R^c}+y_\mu \phi _{T1}\tilde \phi _{\mu _R^c}+y_\tau \phi _{T2}\tilde \phi _{\tau _R^c})h_d/\Lambda \right |^2 \nonumber \\
&+\left |(y_e\phi _{T2}\tilde \phi _{e_R^c}+y_\mu \phi _{T3}\tilde \phi _{\mu _R^c}+y_\tau \phi _{T1}\tilde \phi _{\tau _R^c})h_d/\Lambda \right |^2 \nonumber \\
&+\left |y_e(\phi _{T1}\tilde \phi _{\ell_1}+\phi _{T2}\tilde \phi _{\ell _3}+\phi _{T3}\tilde \phi _{\ell _2})h_d/\Lambda \right |^2 \nonumber \\
&+\left |y_\mu (\phi _{T3}\tilde \phi _{\ell _3}+\phi _{T1}\tilde \phi _{\ell _2}+\phi _{T2}\tilde \phi _{\ell _1})h_d/\Lambda \right |^2 \nonumber \\
&+\left |y_\tau (\phi _{T2}\tilde \phi _{\ell _2}+\phi _{T3}\tilde \phi _{\ell _1}+\phi _{T1}\tilde \phi _{\ell _3})h_d/\Lambda \right |^2 \nonumber \\
&+\left |\left [y_e(\phi _{T1}\tilde \phi _{\ell _1}+\phi _{T2}\tilde \phi _{\ell _3}+\phi _{T3}\tilde \phi _{\ell _2})\tilde \phi _{e_R^c}+y_\mu (\phi _{T3}\tilde \phi _{\ell _3}+\phi _{T1}\tilde \phi _{\ell _2}+\phi _{T2}\tilde \phi _{\ell _1})\tilde \phi _{\mu _R^c}\right .\right . \nonumber \\
&+\left . \left . y_\tau (\phi _{T2}\tilde \phi _{\ell _2}+\phi _{T3}\tilde \phi _{\ell _1}+\phi _{T1}\tilde \phi _{\ell _3})\tilde \phi _{\tau _R^c}\right ]/\Lambda \right |^2,
\end{align}
where $Y=\phi _{T1}$, $\phi _{T2}$, $\phi _{T3}$, $\tilde \phi _{\ell _1}$, $\tilde \phi _{\ell _2}$, $\tilde \phi _{\ell _3}$, $\tilde \phi _{e_R^c}$, $\tilde \phi _{\mu _R^c}$, $\tilde \phi _{\tau _R^c}$, and $h_d$ with $\tilde \phi$ scalar fields being slepton ones. 
The F-term contribution for our model is calculated as 
\begin{align}
\int d^2\theta w_{\Phi _T}&=y_e(\phi _{T1}\ell _1+\phi _{T2}\ell _3+\phi _{T3}\ell _2)e_R^ch_d/\Lambda \nonumber \\
&+y_\mu (\phi _{T3}\ell _3+\phi _{T1}\ell _2+\phi _{T2}\ell _1)\mu _R^ch_d/\Lambda \nonumber \\
&+y_\tau (\phi _{T2}\ell _2+\phi _{T3}\ell _1+\phi _{T1}\ell _3)\tau _R^ch_d/\Lambda \nonumber \\
&+y_e(\tilde \psi _{\phi _{T1}}\tilde \phi _{\ell _1}+\tilde \psi _{\phi _{T2}}\tilde \phi _{\ell _3}+\tilde \psi _{\phi _{T3}}\tilde \phi _{\ell _2})e_R^ch_d/\Lambda \nonumber \\
&+y_\mu (\tilde \psi _{\phi _{T3}}\tilde \phi _{\ell _3}+\tilde \psi _{\phi _{T1}}\tilde \phi _{\ell _2}+\tilde \psi _{\phi _{T2}}\tilde \phi _{\ell _1})\mu _R^ch_d/\Lambda \nonumber \\
&+y_\tau (\tilde \psi _{\phi _{T2}}\tilde \phi _{\ell _2}+\tilde \psi _{\phi _{T3}}\tilde \phi _{\ell _1}+\tilde \psi _{\phi _{T1}}\tilde \phi _{\ell _3})\tau _R^ch_d/\Lambda \nonumber \\
&+y_e(\tilde \psi _{\phi _{T1}}\ell _1+\psi _{\phi _{T2}}\ell _3+\psi _{\phi _{T3}}\ell _2)\tilde \phi _{e_R^c}h_d/\Lambda \nonumber \\
&+y_\mu (\tilde \psi _{\phi _{T3}}\ell _3+\psi _{\phi _{T1}}\ell _2+\psi _{\phi _{T2}}\ell _1)\tilde \phi _{\mu _R^c}h_d/\Lambda \nonumber \\
&+y_\tau (\tilde \psi _{\phi _{T2}}\ell _2+\psi _{\phi _{T3}}\ell _1+\psi _{\phi _{T1}}\ell _3)\tilde \phi _{\tau _R^c}h_d/\Lambda \nonumber \\
&+y_e(\tilde \psi _{\phi _T1}\tilde \phi _{\ell _1}+\tilde \psi _{\phi _T2}\tilde \phi _{\ell _3}+\tilde \psi _{\phi _T3}\tilde \phi _{\ell _2})\tilde \phi _{e_R^c}\tilde \psi _{h_d}/\Lambda \nonumber \\
&+y_\mu (\tilde \psi _{\phi _T3}\tilde \phi _{\ell _3}+\tilde \psi _{\phi _T1}\tilde \phi _{\ell _2}+\tilde \psi _{\phi _T2}\tilde \phi _{\ell _1})\tilde \phi _{\mu _R^c}\tilde \psi _{h_d}/\Lambda \nonumber \\
&+y_\tau (\tilde \psi _{\phi _T2}\tilde \phi _{\ell _2}+\tilde \psi _{\phi _T3}\tilde \phi _{\ell _1}+\tilde \psi _{\phi _T1}\tilde \phi _{\ell _3})\tilde \phi _{\tau _R^c}\tilde \psi _{h_d}/\Lambda \nonumber \\
&+y_e(\phi _{T1}\ell _1+\phi _{T2}\ell _3+\phi _{T3}\ell _2)\tilde \phi _{e_R^c}\tilde \psi _{h_d}/\Lambda \nonumber \\
&+y_\mu (\phi _{T3}\ell _3+\phi _{T1}\ell _2+\phi _{T2}\ell _1)\tilde \phi _{\mu _R^c}\tilde \psi _{h_d}/\Lambda \nonumber \\
&+y_\tau (\phi _{T2}\ell _2+\phi _{T3}\ell _1+\phi _{T1}\ell _3)\tilde \phi _{\tau _R^c}\tilde \psi _{h_d}/\Lambda \nonumber \\
&+y_e(\phi _{T1}\tilde \phi _{\ell _1}+\phi _{T2}\tilde \phi _{\ell _3}+\phi _{T3}\tilde \phi _{\ell _2})e_R^c\tilde \psi _{h_d}/\Lambda \nonumber \\
&+y_\mu (\phi _{T3}\tilde \phi _{\ell _3}+\phi _{T1}\tilde \phi _{\ell _2}+\phi _{T2}\tilde \phi _{\ell _1})\mu _R^c\tilde \psi _{h_d}/\Lambda \nonumber \\
&+y_\tau (\phi _{T2}\tilde \phi _{\ell _2}+\phi _{T3}\tilde \phi _{\ell _1}+\phi _{T1}\tilde \phi _{\ell _3})\tau _R^c\tilde \psi _{h_d}/\Lambda \nonumber \\
&-M(\tilde \psi _{\phi _{01}^T}\tilde \psi _{\phi _{T1}}+\tilde \psi _{\phi _{02}^T}\tilde \psi _{\phi _{T3}}+\tilde \psi _{\phi _{03}^T}\tilde \psi _{\phi _{T2}}) \nonumber \\
&+\frac{2}{3}g\left [\phi _{01}^T(\tilde \psi _{\phi _{T1}}\tilde \psi _{\phi _{T1}}-\tilde \psi _{\Phi _{T2}}\tilde \psi _{\phi _{T3}})+\phi _{02}^T(\tilde \psi _{\phi _{T2}}\tilde \psi _{\phi _{T2}}-\tilde \psi _{\phi _{T3}}\tilde \psi _{\phi _{T1}})\right . \nonumber \\
&+\phi _{03}^T(\tilde \psi _{\phi _{T3}}\tilde \psi _{\phi _{T3}}-\tilde \psi _{\phi _{T1}}\tilde \psi _{\phi _{T2}}) \nonumber \\
&+\tilde \psi _{\phi _{01}^T}(\phi _{T1}\tilde \psi _{\phi _{T1}}-\phi _{T2}\tilde \psi _{\phi _{T3}})+\tilde \psi _{\phi _{02}^T}(\phi _{T2}\tilde \psi _{\phi _{T2}}-\phi _{T3}\tilde \psi _{\phi _{T1}})+\tilde \psi _{\phi _{03}^T}(\phi _{T3}\tilde \psi _{\phi _{T3}}-\phi _{T1}\tilde \psi _{\phi _{T2}})\nonumber \\
&+\left .\tilde \psi _{\phi _{01}^T}(\tilde \psi _{\phi _{T1}}\phi _{T1}-\tilde \psi _{\phi _{T2}}\phi _{T3})+
\tilde \psi _{\phi _{02}^T}(\tilde \psi _{\phi _{T2}}\phi _{T2}-\tilde \psi _{\phi _{T3}}\phi _{T1})+\tilde \psi _{\phi _{03}^T}(\tilde \psi _{\phi _{T3}}\phi _{T3}-\tilde \psi _{\phi _{T1}}\phi _{T2})\right ],
\end{align}
where $\tilde \psi$ fields correspond to flavinos.
The VEV alignments of $\phi_T$ and $\phi _0^T$ are obtained from the potential minimum condition, $V_T =0$, as in Ref.~\cite{Morozumi:2017rrg};
\begin{equation}
\langle \phi_T \rangle = v_T (1, 0, 0), \qquad  v_T = \frac{3M}{2g}, \qquad \langle \phi _0^T\rangle =(0,0,0).
\end{equation} 
The charged lepton masses are then obtained after $h_d$ developing its VEV, $\langle h_d \rangle = v_d$, as $m_{\{e,\mu,\tau\}} = y_{\{e,\mu,\tau \}} v_d v_T/\Lambda$. 
After taking the VEV alignments, e.g. the $A_4$ triplet flavon 
($\phi _{T1}$, $\phi _{T2}$, $\phi _{T3}$)=($v_T+\varphi _{T1}$, $\phi _{T2}$, $\phi _{T3}$), 
we find the mass terms of flavons and flavinos as follows,
\begin{align}
\mathcal{L}_{\Phi _T} \supset \ & M^2 |\varphi_{T1}|^2 + 4 M^2  |\phi_{T2}|^2 + 4 M^2  |\phi_{T3}|^2 
+ M^2 |\phi_{01}^T|^2 + 4 M^2  |\phi_{02}^T|^2 + 4 M^2  |\phi_{03}^T|^2  \nonumber \\
&- M \tilde{\psi}_{\phi^T_{01}} \tilde{\psi}_{\phi_{T1}} - 2M (\tilde \psi _{\phi _{02}^T}\tilde \psi _{\phi _{T3}}+\tilde \psi _{\phi _{03}^T}\tilde \psi _{\phi _{T2}}) + h.c. \, .
\end{align}
Notice that pairs of chiral fermion components $\{\tilde{\psi}_{\phi^T_{01}},  \tilde{\psi}_{\phi_{T1}}\}$, $\{\tilde \psi _{\phi _{02}^T},  \tilde \psi _{\phi _{T3}}\}$ and 
$\{\tilde \psi _{\phi _{03}^T},  \tilde \psi _{\phi _{T2}}\}$ construct Dirac mass terms.
We also find that the mass ratio of flavons(flavinos) is $1:2:2$, and flavino and flavon masses are degenerate without considering SUSY breaking effects.
In this work the DM candidate is the lightest flavino which is Dirac fermion with mass $M$. 
We rewrite it as 
\begin{equation}
X_R \equiv \tilde{\psi}^c_{\phi^T_{01}}, \qquad X_L \equiv \tilde{\psi}_{\phi_{T1}},
\end{equation} 
where $"c"$ indicates charge conjugation and $X$ is identified as Dirac fermion.
Here we focus on interactions among the DM $X$ and the lightest flavons $\varphi_{T1}$ and $\phi^T_{01}$ since it provide dominant contributions in the relic density estimation as we discuss in next section.
The relevant interactions are summarized by
\begin{align}
\mathcal{L}_{\Phi _T} \supset \  &  \frac{M}{v_T} \left[ 2 \varphi_{T1} \overline{X_R} X_L + \phi_{01}^T \overline{X^c_L} X_L + h.c.  \right] \nonumber \\
& \frac{M^2}{v_T} \left[ \varphi_{T1}  \varphi_{T1}^* \varphi_{T1}^* + c.c. \right] - \frac{2M^2}{v_T} \left[ \phi_{01}^T  \phi_{01}^{T*} \varphi_{T1}^* + c.c. \right].
\end{align}
These interactions induce DM annihilation processes that determine the relic density.


\section{Dark matter physics}
\label{sec:Dark-matter}

In this section we discuss the relic density of flavino DM calculating its annihilation cross section.
Other DM phenomenologies such as direct/indirect detections and collider searches are also discussed briefly.

\subsection{DM annihilation cross sections}

In our scenario the flavinos mainly annihilate into the lightest flavons, $\{\varphi_{T},  \phi_{01}^T\}$, due to small couplings among flavino and the SM leptons. 
There are many possible processes of such annihilation processes as
\begin{align}
& \overline{X} X \rightarrow \{ \varphi_{T1} \varphi_{T1},  \overline{\varphi_{T1}} \ \overline{\varphi_{T1}},  \overline{\phi^T_{01}} \phi^T_{01} ,   \overline{\varphi_{T1}} \varphi_{T1}\},  \nonumber \\
& X X \rightarrow \overline{\phi^T_{01}} \ \overline{\varphi_{T1}}, \quad \overline{X} \ \overline{X} \rightarrow \phi^T_{01} \varphi_{T1}.
\end{align}
Although many diagrams can be drawn from the Lagrangian, we find the dominant contributions are $\overline{X} X \rightarrow \{\varphi_{T1} \varphi_{T1}, \overline{\varphi_{T1}} \ \overline{\varphi_{T1}} \}$ and $X X \rightarrow \overline{\phi^T_{01}} \ \overline{\varphi_{T1}} (\overline{X} \ \overline{X} \rightarrow \phi^T_{01} \varphi_{T_1})$ where the other processes are suppressed by $p^2/M_{}^2$ factor with $p$ being magnitude of DM momentum.
In the following we summarize the cross sections of the dominant processes.
\\

\noindent
(i) $X\bar{X} \longrightarrow \varphi_{T1} \varphi_{T1}$: \\
In this process, all channels are present and
relevant matrix elements are
\begin{figure}[H]
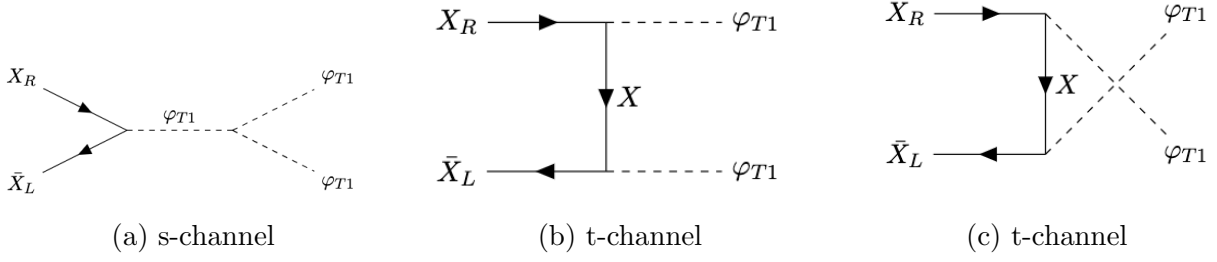

  \centering
  \begin{minipage}[b]{0.32\linewidth}
    \centering
    \includegraphics[width=0.9\columnwidth]{phiphi_s-cha.pdf}
    \label{fig:s-cha}
    \subcaption{s-channel}
  \end{minipage}
  \begin{minipage}[b]{0.32\linewidth}
    \centering
    \includegraphics[width=0.9\columnwidth]{phiphi_t-cha.pdf}
    \label{fig:t-cha}
    \subcaption{t-channel}
  \end{minipage}
  \begin{minipage}[b]{0.32\columnwidth}
    \centering
    \includegraphics[width=0.9\columnwidth]{phiphi_u-cha.pdf}
    \label{fig:u-cha}
    \subcaption{t-channel}
\end{minipage}
  \caption{Diagrams for the process $\bar X X \to \varphi_{T1} \varphi_{T1}$.}
 \end{figure}
\begin{align}
  \mathcal{M}_s &= -\left(\frac{2M}{v_T}\right) \left(\frac{2M^2}{v_T}\right) \frac{1}{s-M^2}\bar{v}_{(p_2)} P_R u_{(p_1)}\notag \\
  \mathcal{M}_t &\thicksim {\left(\frac{M}{v_T}\right)}^2 \frac{1}{M} \bar{v}_{(p_2)} P_R u_{(p_1)}\notag \\
  \mathcal{M}_u &\thicksim {\left(\frac{M}{v_T}\right)}^2 \frac{1}{M} \bar{v}_{(p_2)} P_R u_{(p_1)},
\end{align} 
where $s = (p_1 + p_2)^2$ with $p_{1(2)}$ being flavino $X(\bar X)$ momentum in the initial state.
Here, we ignored other Mandelstam variables assuming $t,u \ll M^2 $.
Since we consider non-retivistic limit, we approximate $s\thicksim 4M^2 $.
Thus, the matrix element of the process is
\begin{align}
  \mathcal{M} \thicksim \frac{20M}{3{v_T}^2} \bar{v}_{(p_2)} P_R u_{(p_1)}.\notag 
\end{align}
The cross section is
\begin{align}
  \sigma_{X\bar{X} \rightarrow  \varphi_{T1} \varphi_{T1}}  \thicksim \frac{25}{72\pi} \frac{M^2}{{v_T}^4} \frac{s-2M^2}{s}.
\end{align}

\noindent
(ii)$X\bar{X} \longrightarrow \overline{\varphi_{T1}} \, \overline{\varphi_{T1}} $: \\
This process are just conjugate of $X\bar{X} \longrightarrow \varphi_{T1} \varphi_{T1}$.
Thus, cross section is
\begin{align}
  \sigma_{X\bar{X} \rightarrow \overline{\varphi_{T1}} \, \overline{\varphi_{T1}} } &= \sigma_{X\bar{X} \rightarrow  \varphi_{T1} \varphi_{T1}}\notag \\
                   &\thicksim \frac{25}{72\pi} \frac{M^2}{{v_T}^4} \frac{s-2M^2}{s}.
\end{align}

\noindent
(iii)$XX \longrightarrow \overline{\phi^{T}_{01}} \, \overline{{\phi}_{T1}}$: \\
In this prosess, all channels are present.
Matrix elements are
\begin{figure}[H]
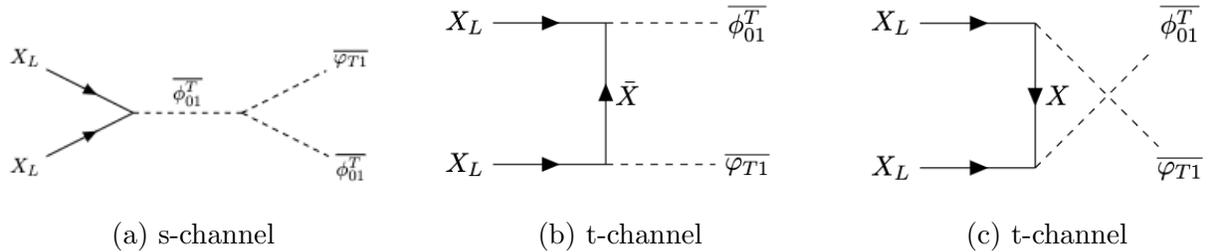

  \centering
  \begin{minipage}[b]{0.32\linewidth}
    \centering
    \includegraphics[width=0.9\columnwidth]{barphiphi_s-cha.pdf}
    \label{fig:s-cha}
    \subcaption{s-channel}
  \end{minipage}
  \begin{minipage}[b]{0.32\linewidth}
    \centering
    \includegraphics[width=0.9\columnwidth]{barphiphi_t-cha.pdf}
    \label{fig:t-cha}
    \subcaption{t-channel}
  \end{minipage}
  \begin{minipage}[b]{0.32\columnwidth}
    \centering
    \includegraphics[width=0.9\columnwidth]{barphiphi_u-cha.pdf}
    \label{fig:u-cha}
    \subcaption{t-channel}
\end{minipage}
  \caption{Diagrams for the process $XX \to \overline{\varphi_{T1}} \overline{\phi^T_{01}}$. }
 \end{figure}
\begin{align}
  \mathcal{M}_s &= \left(\frac{M}{v_T}\right) \left(\frac{2M^2}{v_T}\right) \frac{1}{s-M^2}\bar{v}_{(p_2)} P_L u_{(p_1)}\notag \\
  \mathcal{M}_t &\thicksim  \mathcal{M}_u \thicksim \left(\frac{2M}{{v_T}^2}\right) \bar{v}_{(p_2)} P_L u_{(p_1)} 
\end{align}
This process's matrix element is
\begin{align}
  \mathcal{M} \thicksim \left(\frac{14M}{{3v_T}^2}\right) \bar{v}_{(p_2)} P_L u_{(p_1)}.\notag
\end{align}
The cross section is 
\begin{align}
  \sigma_{XX \rightarrow \overline{\phi^{T}_{01}}  \overline{\phi_{T1}} } \thicksim \left(\frac{49}{144\pi s}\right) \left(\frac{{M}^2}{{v_T}^4}\right) \left( s-2M^2 \right).
\end{align}

\noindent
(iv)$\bar{X} \bar{X} \longrightarrow {\phi}^{T}_{01} {\varphi}_{T1}$: \\
This process are just conjugate of $XX \longrightarrow  \overline{\phi^{T}_{01}} \overline{\varphi_{T1}}$. 
Thus, cross section is
\begin{align}
  \sigma_{\bar{X} \bar{X} \rightarrow {\phi}^{T}_{01} {\varphi}_{T1}} &=  \sigma_{XX \rightarrow \overline{\varphi^{T}_{01}} \overline{\varphi_{T1}} } \nonumber \\
  &\thicksim \left(\frac{49}{144\pi s}\right) \left(\frac{{M}^2}{{v_T}^4}\right) \left( s-2M^2 \right).
\end{align}
We apply these cross sections in DM relic density calculation.

\subsection{Relic density}

Relic density of flavino DM is obtained by solving Boltzmann equation for number density $n_X$ of DM $X$,
\begin{equation}
\dot{n}_X + 3 H n_X = \langle \sigma v \rangle (n_{X_{\rm eq}}^2 - n_X^2), 
\end{equation}
where $H$ is Hubble parameter, $\langle \sigma v \rangle$ is thermal average of DM annihilation cross section and $n_{X_{eq}}$ is density of $X$ in equilibrium.
The annihilation cross section is the sum of cross sections discussed in the previous subsection.
Relic density $\Omega_X h^2$ can be approximately estimated as~\cite{Griest:1990kh, Gondolo:1990dk}
\begin{align}
\Omega_X h^2 \simeq \frac{1.07 \times 10^9}{\sqrt{g^*(x_f)} M_{Pl} J(x_f) \ {\rm [GeV]}},
\label{eq:relic}
\end{align}
where $x_f = M/T_f$ with $T_f$ being freeze out temperature, $g^*(x_f)$ is effective relativistic degrees of freedom at $T_f$, and $M_{Pl} \simeq 1.22 \times 10^{19}$ is the Planck mass.
The factor $J(x_f) \equiv \int_{x_f}^\infty dx \frac{\langle \sigma v \rangle}{x^2}$ is written by
\begin{equation}
J(x_f ) = \int_{x_f}^{\infty} \left[ \frac{\int_{4 M^2}^{\infty} ds \sqrt{s - 4 M^2} (\sigma v) K_1 \left( \frac{\sqrt{s} }{M} x \right) }{16 M^5 x [K_2(x)]^2} \right],
\end{equation}
where $K_{1,2}$ denote the modified Bessel functions of the second kind of order 1 and 2.

 \begin{figure}[tb]
 \begin{center}
\includegraphics[width=8cm]{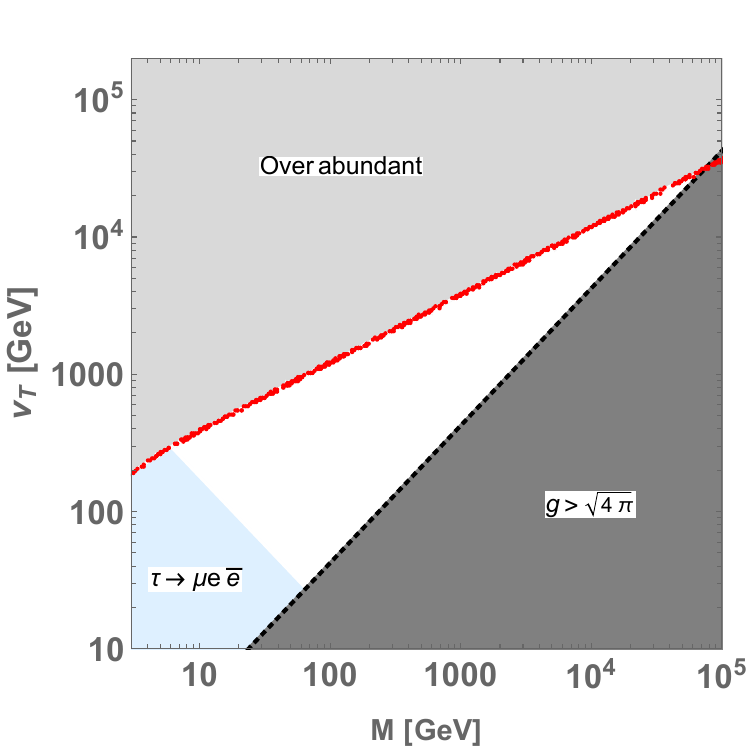} \ 
\caption{Red points shows the parameters that provide observed relic density $\Omega_X h^2 \simeq 0.12$. The gray region is disfavored since coupling $g$ is beyond the perturbative limit. Relic density is over abundant in the light gray region. The light blue region is excluded by the constraint from flavor violating $\tau$ decay. }
\label{fig:DM}
\end{center}
\end{figure}

In estimating relic density we apply {\it micrOMEGAs} code~\cite{Belanger:2014vza} implementing relevant interactions in the model; we compared the numerical results with analytical calculation with Eq.~\eqref{eq:relic} 
and confirmed consistency.
In Fig.~\ref{fig:DM}, we show parameters that provide observed amount of relic density $\Omega_X h^2 \simeq 0.12$~\cite{Planck:2018vyg} by red points where relic density is over abundant in the light gray region 
and smaller than observed amount below red points. Also the gray region is disfavored since coupling $g$ becomes non-perturbative value. 
In addition, the light blue region is excluded by the constraint from flavor violating $\tau$ decay~\cite{Muramatsu:2016bda} (see Appendix \ref{sec:LFVconstraint}).
We thus find that the range of the lightest flavino(flavon) mass $[6, 6 \times 10^4]$ [GeV] can accommodate observed relic density.
Also allowed region of $v_T$ is roughly 30 [GeV] to $2 \times 10^4$ [GeV].

\subsection{Comments on ohter flavino DM phyiscs}

Here we provide brief discussion regarding some phenomenology of flavino DM.
\\

\noindent
{\it Direct detection}:\\
The interactions from superpotential do not contribute to flavino-nucleon scattering at tree and one-loop level since they only have coupling terms among flavino, flavon, Higgs(Higgsinos) and leptons(sleptons); there is no flavino-flavino-Higgs interaction.
Thus these interactions are not constrained by the direct detection experiments; flavino can also interact with electron via flavon $\varphi_{T1}$ exchange but cross section is tiny since flavon-electron coupling is proportional to electron mass.
In fact flavino-nucleon scattering is possible via Higgs portal interactions when flavon and Higgs bosons mix through SUSY-breaking terms. 
In this work we assume effect of SUSY breaking is small for flavon sector to avoid current experimental bounds, e.g. XENON1T~\cite{XENON:2018voc}, PandaX-4T~\cite{PandaX-4T:2021bab} and LUX-ZEPLIN~\cite{LZ:2022lsv}.
\\

\noindent
{\it Indirect detection}:\\
In the scenario cross sections of flavino annihilation into flavons are suppressed at current universe due to mass degeneracy between the lightest flavino and flavon.
Flavino can also annihilate into charged leptons via flavon exchanging processes, $X \bar{X} \to \varphi_{T1} \to \ell^+ \ell^-$, but cross section of the processes 
are much smaller than order of $\sim 10^{-26}$cm$^2/$s, since flavon-lepton coupling is small as $m_\ell/v_\varphi$.
Thus the model is safe from indirect detection constraints where the strongest bound on the annihilation cross section is given by Fermi-LAT data~\cite{Fermi-LAT:2015att, Circiello:2024gpq}.
\\

\noindent
{\it Collider search}: \\
Flavino DM can be searched for at collider experiments such as the LHC as the preferred scale of flavino mass is $\mathcal{O}$(10) GeV to $\mathcal{O}$(10) TeV.
One possible process is slepton pair production followed by slepton decay $\tilde \ell \to \ell X$ when a slepton is the next to lightest SUSY particle.
The signal of this case is the same as slepton decaying into neutralino DM~\cite{Roszkowski:2017nbc} and it is difficult to distinguish.
We can expect more specific signals of the model when heavier flavinos and flavons are lighter than sleptons. 
In such a case we would have cascade decay of slepton, e.g. $\tilde \ell \to  \ell \overline{\tilde{\psi}_{T2}}  \to \ell \overline{X} \phi_{T3} (\to \bar{\ell}' \ell') $, 
inducing multi-leptons with missing-energy signal.
For these signals we may be able to reconstruct flavon mass and it helps us to confirm the model.
Detailed analysis of collider signals is beyond the scope of this paper and we left it in future work.



\section{Summary}
\label{sec:Summary}
We have studied the relic density of the flavino DM in modified AF $A_4$ model 
which is respecting the $SU(2)_L\times A_4\times Z_3\times U(1)_R$ symmetry. 
First of all, we have calculated the Lagrangian from the superpotential in the modified AF $A_4$ model.
In estimating relic density, we have considered the relevant interactions from the Lagrangian 
that realize the VEV alignments and charged lepton masses where we assume SUSY breaking effects 
are small for flavon sector.  
As a result, we have found the degenerate masses between the lightest flavon and flavino, 
and only two parameters in the potential determines the relic density.
Then allowed parameter space of these parameters have been estimated from relic density calculation taking a 
constraint from lepton flavor violation (LFV) into account. 
We thus have found that the range of lightest flavino(flavon) mass $[6, 6 \times 10^4]$ [GeV] could 
accommodate observed relic density.
Also allowed region of $v_T$ has been roughly 30 [GeV] to $2 \times 10^4$ [GeV].  
We also briefly have discussed other DM physics such as direct detection, indirect detection and collider search.

\newpage 
\vspace{1cm}
\noindent
{\large \bf Acknowledgement}
\vspace{1mm}

We thank T. Toma for useful discussions. 
The work of T.~N. was supported by the Fundamental Research Funds for the Central Universities.
The work of Y.~S. was partially supported by JSPS KAKENHI Grant No. 20H01898.

\appendix
\section*{Appendix}

\section{Multiplication rule of $A_4$ group}
\label{sec:multiplication-rule}
We use the multiplication rule of the $A_4$ triplet as follows:
\begin{align}
\label{eq:multiplication-rule}
\begin{pmatrix}
a_1\\
a_2\\
a_3
\end{pmatrix}_{\bf 3}
\otimes
\begin{pmatrix}
b_1\\
b_2\\
b_3
\end{pmatrix}_{\bf 3}
&=\left (a_1b_1+a_2b_3+a_3b_2\right )_{\bf 1}
\oplus \left (a_3b_3+a_1b_2+a_2b_1\right )_{{\bf 1}'} \nonumber \\
& \oplus \left (a_2b_2+a_1b_3+a_3b_1\right )_{{\bf 1}''} \nonumber \\
&\oplus \frac13
\begin{pmatrix}
2a_1b_1-a_2b_3-a_3b_2 \\
2a_3b_3-a_1b_2-a_2b_1 \\
2a_2b_2-a_3b_1-a_1b_3
\end{pmatrix}_{{\bf 3}}
\oplus \frac12
\begin{pmatrix}
a_2b_3-a_3b_2 \\
a_1b_2-a_2b_1 \\
a_3b_1-a_1b_3
\end{pmatrix}_{{\bf 3}\  } \ .
\end{align}
More details are shown in the review~\cite{Ishimori:2010au,Ishimori:2012zz}.


\section{LFV constraint in the model}
\label{sec:LFVconstraint}

In the model some LFV charged lepton decay processes are forbidden by residual $Z_3$ symmetry~\cite{Muramatsu:2016bda}. 
In particular LFV muon decays are forbidden and the strongest constraint is obtained from $\tau \to \mu \mu \bar{e}$ process. 
The branching ratio of the process is given by
\begin{align}
{\rm BR}(\tau \to \mu \mu \bar{e}) &= \tau_\tau \frac{m_\tau^5}{3027 \pi^3} \left( \left| \frac{m_\tau m_\mu}{v_T^2 m^2_{\phi_{T2}}} \right|^2 + \left| \frac{m_\mu m_e}{v_T^2 m^2_{\phi_{T3}}} \right|^2 \right)
\nonumber \\
\simeq & \frac{2.9 \times 10^6 \, {\rm GeV}^8}{v_T^4 (2 M)^4}
\end{align}
where $\tau_\tau$ is the lifetime of tau lepton.
It is compared with the current constraint by measurement at Belle experiment BR$(\tau \to \mu \mu \bar{e}) < 1.7 \times 10^{-8}$~\cite{Hayasaka:2010np} to constrain $\{M, v_T \}$ parameter space.



\begin{thebibliography}{99}

\bibitem{DayaBay:2012fng}
F.~P.~An \textit{et al.} [Daya Bay],
Phys. Rev. Lett. \textbf{108} (2012), 171803
[arXiv:1203.1669 [hep-ex]].

\bibitem{RENO:2012mkc}
J.~K.~Ahn \textit{et al.} [RENO],
Phys. Rev. Lett. \textbf{108} (2012), 191802
[arXiv:1204.0626 [hep-ex]].


\bibitem{Froggatt:1978nt}
C.~D.~Froggatt and H.~B.~Nielsen,
Nucl. Phys. B \textbf{147} (1979), 277-298




\bibitem{Altarelli:2010gt}
G.~Altarelli and F.~Feruglio,
Rev. Mod. Phys. \textbf{82} (2010), 2701-2729
[arXiv:1002.0211 [hep-ph]].


\bibitem{Ishimori:2010au}
H.~Ishimori, T.~Kobayashi, H.~Ohki, Y.~Shimizu, H.~Okada and M.~Tanimoto,
Prog. Theor. Phys. Suppl. \textbf{183} (2010), 1-163
[arXiv:1003.3552 [hep-th]].


\bibitem{Ishimori:2012zz}
H.~Ishimori, T.~Kobayashi, H.~Ohki, H.~Okada, Y.~Shimizu and M.~Tanimoto,
Lect. Notes Phys. \textbf{858} (2012), 1-227, Springer.


\bibitem{King:2014nza}
S.~F.~King, A.~Merle, S.~Morisi, Y.~Shimizu and M.~Tanimoto,
New J. Phys. \textbf{16} (2014), 045018
[arXiv:1402.4271 [hep-ph]].


\bibitem{Kobayashi:2022moq}
T.~Kobayashi, H.~Ohki, H.~Okada, Y.~Shimizu and M.~Tanimoto,
Lect. Notes Phys. \textbf{995} (2022), 1-353, Springer.




\bibitem{Harrison:2002er}
P.~F.~Harrison, D.~H.~Perkins and W.~G.~Scott,
Phys. Lett. B \textbf{530} (2002), 167
[arXiv:hep-ph/0202074 [hep-ph]].

\bibitem{Harrison:2002kp}
P.~F.~Harrison and W.~G.~Scott,
Phys. Lett. B \textbf{535} (2002), 163-169
[arXiv:hep-ph/0203209 [hep-ph]].






\bibitem{Xing:2002sw}
Z.~z.~Xing,
Phys. Lett. B \textbf{533} (2002), 85-93
[arXiv:hep-ph/0204049 [hep-ph]].

\bibitem{Xing:2006ms}
Z.~z.~Xing and S.~Zhou,
Phys. Lett. B \textbf{653} (2007), 278-287
[arXiv:hep-ph/0607302 [hep-ph]].


\bibitem{Adhikary:2006jx}
B.~Adhikary and A.~Ghosal,
Phys. Rev. D \textbf{75} (2007), 073020
[arXiv:hep-ph/0609193 [hep-ph]].

\bibitem{King:2007pr}
S.~F.~King,
Phys. Lett. B \textbf{659} (2008), 244-251
[arXiv:0710.0530 [hep-ph]].

\bibitem{Honda:2008rs}
M.~Honda and M.~Tanimoto,
Prog. Theor. Phys. \textbf{119} (2008), 583-598
[arXiv:0801.0181 [hep-ph]].

\bibitem{Brahmachari:2008fn}
B.~Brahmachari, S.~Choubey and M.~Mitra,
Phys. Rev. D \textbf{77} (2008), 073008
[erratum: Phys. Rev. D \textbf{77} (2008), 119901]
[arXiv:0801.3554 [hep-ph]].

\bibitem{Adhikary:2008au}
B.~Adhikary and A.~Ghosal,
Phys. Rev. D \textbf{78} (2008), 073007
[arXiv:0803.3582 [hep-ph]].

\bibitem{Hirsch:2008mg}
M.~Hirsch, S.~Morisi and J.~W.~F.~Valle,
Phys. Rev. D \textbf{79} (2009), 016001
[arXiv:0810.0121 [hep-ph]].

\bibitem{Morisi:2009qa}
S.~Morisi,
Phys. Rev. D \textbf{79} (2009), 033008
[arXiv:0901.1080 [hep-ph]].

\bibitem{Hayakawa:2009va}
A.~Hayakawa, H.~Ishimori, Y.~Shimizu and M.~Tanimoto,
Phys. Lett. B \textbf{680} (2009), 334-342
[arXiv:0904.3820 [hep-ph]].

\bibitem{Goswami:2009yy}
S.~Goswami, S.~T.~Petcov, S.~Ray and W.~Rodejohann,
Phys. Rev. D \textbf{80} (2009), 053013
[arXiv:0907.2869 [hep-ph]].

\bibitem{Barry:2010zk}
J.~Barry and W.~Rodejohann,
Phys. Rev. D \textbf{81} (2010), 093002
[erratum: Phys. Rev. D \textbf{81} (2010), 119901]
[arXiv:1003.2385 [hep-ph]].

\bibitem{Albright:2010ap}
C.~H.~Albright, A.~Dueck and W.~Rodejohann,
Eur. Phys. J. C \textbf{70} (2010), 1099-1110
[arXiv:1004.2798 [hep-ph]].

\bibitem{Ishimori:2010fs}
H.~Ishimori, Y.~Shimizu, M.~Tanimoto and A.~Watanabe,
Phys. Rev. D \textbf{83} (2011), 033004
[arXiv:1010.3805 [hep-ph]].

\bibitem{King:2011zj}
S.~F.~King and C.~Luhn,
JHEP \textbf{09} (2011), 042
[arXiv:1107.5332 [hep-ph]].

\bibitem{King:2011ab}
S.~F.~King and C.~Luhn,
JHEP \textbf{03} (2012), 036
[arXiv:1112.1959 [hep-ph]].

\bibitem{Shimizu:2011xg}
Y.~Shimizu, M.~Tanimoto and A.~Watanabe,
Prog. Theor. Phys. \textbf{126} (2011), 81-90
[arXiv:1105.2929 [hep-ph]].

\bibitem{Antusch:2011qg}
S.~Antusch and V.~Maurer,
Phys. Rev. D \textbf{84} (2011), 117301
[arXiv:1107.3728 [hep-ph]].


\bibitem{Ahn:2012tv}
Y.~H.~Ahn and S.~K.~Kang,
Phys. Rev. D \textbf{86} (2012), 093003
[arXiv:1203.4185 [hep-ph]].

\bibitem{Ishimori:2012fg}
H.~Ishimori and E.~Ma,
Phys. Rev. D \textbf{86} (2012), 045030
[arXiv:1205.0075 [hep-ph]].

\bibitem{Rodejohann:2012cf}
W.~Rodejohann and H.~Zhang,
Phys. Rev. D \textbf{86} (2012), 093008
[arXiv:1207.1225 [hep-ph]].

\bibitem{Hagedorn:2012ut}
C.~Hagedorn, S.~F.~King and C.~Luhn,
Phys. Lett. B \textbf{717} (2012), 207-213
[arXiv:1205.3114 [hep-ph]].

\bibitem{King:2013eh}
S.~F.~King and C.~Luhn,
Rept. Prog. Phys. \textbf{76} (2013), 056201
[arXiv:1301.1340 [hep-ph]].






\bibitem{King:2012vj}
S.~F.~King,
Phys. Lett. B \textbf{718} (2012), 136-142
[arXiv:1205.0506 [hep-ph]].


\bibitem{Shimizu:2012ry}
Y.~Shimizu, R.~Takahashi and M.~Tanimoto,
PTEP \textbf{2013} (2013) no.6, 063B02
[arXiv:1212.5913 [hep-ph]].





\bibitem{Altarelli:2005yp}
G.~Altarelli and F.~Feruglio,
Nucl. Phys. B \textbf{720} (2005), 64-88
[arXiv:hep-ph/0504165 [hep-ph]].

\bibitem{Altarelli:2005yx}
G.~Altarelli and F.~Feruglio,
Nucl. Phys. B \textbf{741} (2006), 215-235
[arXiv:hep-ph/0512103 [hep-ph]].



\bibitem{Farrar:1978xj}
G.~R.~Farrar and P.~Fayet,
Phys. Lett. B \textbf{76} (1978), 575-579

\bibitem{Ellis:1983ew}
J.~R.~Ellis, J.~S.~Hagelin, D.~V.~Nanopoulos, K.~A.~Olive and M.~Srednicki,
Nucl. Phys. B \textbf{238} (1984), 453-476

\bibitem{Roszkowski:2017nbc}
L.~Roszkowski, E.~M.~Sessolo and S.~Trojanowski,
Rept. Prog. Phys. \textbf{81} (2018) no.6, 066201
[arXiv:1707.06277 [hep-ph]].



\bibitem{Baer:2004qq}
H.~Baer, A.~Belyaev, T.~Krupovnickas and J.~O'Farrill,
JCAP \textbf{08} (2004), 005
[arXiv:hep-ph/0405210 [hep-ph]].

\bibitem{Baer:2005bu}
H.~Baer, A.~Mustafayev, S.~Profumo, A.~Belyaev and X.~Tata,
JHEP \textbf{07} (2005), 065
[arXiv:hep-ph/0504001 [hep-ph]].

\bibitem{Han:2014nba}
T.~Han, Z.~Liu and S.~Su,
JHEP \textbf{08} (2014), 093
[arXiv:1406.1181 [hep-ph]].

\bibitem{GAMBIT:2018gjo}
P.~Athron \textit{et al.} [GAMBIT],
Eur. Phys. J. C \textbf{79} (2019) no.5, 395
[arXiv:1809.02097 [hep-ph]].

\bibitem{ATLAS:2021moa}
G.~Aad \textit{et al.} [ATLAS],
Eur. Phys. J. C \textbf{81} (2021) no.12, 1118
[arXiv:2106.01676 [hep-ex]].

\bibitem{CMS:2021few}
A.~Tumasyan \textit{et al.} [CMS],
JHEP \textbf{10} (2021), 045
[arXiv:2107.12553 [hep-ex]].













\bibitem{Chigusa:2018yua}
S.~Chigusa, S.~Kasuya and K.~Nakayama,
Phys. Lett. B \textbf{788} (2019), 494-499
[arXiv:1810.05791 [hep-ph]].







\bibitem{Muramatsu:2016bda}
Y.~Muramatsu, T.~Nomura and Y.~Shimizu,
JHEP \textbf{03} (2016), 192
[arXiv:1601.04788 [hep-ph]].


\bibitem{Nomura:2016nfi}
T.~Nomura, Y.~Shimizu and T.~Yamada,
JHEP \textbf{06} (2016), 125
[arXiv:1604.07650 [hep-ph]].

\bibitem{Morozumi:2017rrg}
T.~Morozumi, H.~Okane, H.~Sakamoto, Y.~Shimizu, K.~Takagi and H.~Umeeda,
Chin. Phys. C \textbf{42} (2018) no.2, 023102
[arXiv:1707.04028 [hep-ph]].


\bibitem{Muramatsu:2017xmn}
Y.~Muramatsu, T.~Nomura, Y.~Shimizu and H.~Yokoya,
Phys. Rev. D \textbf{97} (2018) no.1, 015003
[arXiv:1707.06542 [hep-ph]].





\bibitem{Griest:1990kh}
K.~Griest and D.~Seckel,
Phys. Rev. D \textbf{43} (1991), 3191-3203

\bibitem{Gondolo:1990dk}
P.~Gondolo and G.~Gelmini,
Nucl. Phys. B \textbf{360} (1991), 145-179


\bibitem{Belanger:2014vza}
G.~B\'elanger, F.~Boudjema, A.~Pukhov and A.~Semenov,
Comput. Phys. Commun. \textbf{192} (2015), 322-329
[arXiv:1407.6129 [hep-ph]].


\bibitem{Planck:2018vyg}
N.~Aghanim \textit{et al.} [Planck],
Astron. Astrophys. \textbf{641} (2020), A6
[erratum: Astron. Astrophys. \textbf{652} (2021), C4]
[arXiv:1807.06209 [astro-ph.CO]].


\bibitem{XENON:2018voc}
E.~Aprile \textit{et al.} [XENON],
Phys. Rev. Lett. \textbf{121} (2018) no.11, 111302
[arXiv:1805.12562 [astro-ph.CO]].

\bibitem{PandaX-4T:2021bab}
Y.~Meng \textit{et al.} [PandaX-4T],
Phys. Rev. Lett. \textbf{127} (2021) no.26, 261802
[arXiv:2107.13438 [hep-ex]].

\bibitem{LZ:2022lsv}
J.~Aalbers \textit{et al.} [LZ],
Phys. Rev. Lett. \textbf{131} (2023) no.4, 041002
[arXiv:2207.03764 [hep-ex]].





\bibitem{Fermi-LAT:2015att}
M.~Ackermann \textit{et al.} [Fermi-LAT],
Phys. Rev. Lett. \textbf{115} (2015) no.23, 231301
[arXiv:1503.02641 [astro-ph.HE]].

\bibitem{Circiello:2024gpq}
A.~Circiello, A.~McDaniel, A.~Drlica-Wagner, C.~Karwin, M.~Ajello, M.~Di Mauro and M.~\'A.~S\'anchez-Conde,
[arXiv:2404.01181 [astro-ph.HE]].



\bibitem{Hayasaka:2010np}
K.~Hayasaka, K.~Inami, Y.~Miyazaki, K.~Arinstein, V.~Aulchenko, T.~Aushev, A.~M.~Bakich, A.~Bay, K.~Belous and V.~Bhardwaj, \textit{et al.}
Phys. Lett. B \textbf{687} (2010), 139-143
[arXiv:1001.3221 [hep-ex]].


\end{thebibliography}
\end{document}